\documentstyle[aps,pre,twocolumn,floats,epsfig,amssymb]{revtex}

\begin{document}

\twocolumn[\hsize\textwidth\columnwidth\hsize\csname @twocolumnfalse\endcsname

\title{Dynamical exponents of an even-parity-conserving
             contact process with diffusion}

\author{J. Ricardo G. de Mendon\c{c}a\cite{email}}

\address{Departamento de Estat\'{\i}stica, Instituto de Matem\'{a}tica e
Estat\'{\i}stica,  Universidade  de  S\~{a}o  Paulo \\ Rua do  Mat\~{a}o
1010, Cidade Universit\'{a}ria, 05508-900 S\~{a}o Paulo, SP, Brazil}

\address{Ericsson  Research  \&\  Development  Center  ---  CDMA Systems
Development  Unit \\ Rodovia  Erm\^{e}nio de Oliveira  Penteado km 57.5,
Bairro do Tombadouro, 13330-000 Indaiatuba, SP, Brazil}

\date{\today}

\maketitle

\begin{abstract}
We provide  finite-size  scaling  estimates for the  dynamical  critical
exponent of the even  parity-conserving  universality  class of critical
behavior through exact numerical  diagonalizations of the time evolution
operator of an even-parity-conserving contact process.  Our data seem to
indicate that upon the  introduction  of a small  diffusion  rate in the
process  its  critical  behavior  crosses  over to that of the  directed
percolation  universality  class.  A brief discussion of the many-sector
decomposability of parity-conserving contact processes is presented.
\end{abstract}

 \vspace{2pc}]

\section{Introduction}

One of the most  important  theoretical  programs in modern  statistical
mechanics  is  that  of the  understanding  and  classification  of  the
critical behavior of  nonequilibrium  interacting  particle  systems.  A
first  major  step  towards  this  program  was  given by the  so-called
directed  percolation (DP) conjecture  \cite{zpb42,zpb47},  which in its
original formulation stated that all dynamically driven continuous phase
transitions about a single absorbing state in  single-component  systems
with a scalar order parameter in the absence of internal  symmetries are
in the DP  universality  class of critical  behavior  \cite{kinzel}.  In
this form the conjecture has been confirmed in a host of model  systems,
among  others the basic  contact  process  \cite{harris},  Schl\"{o}gl's
models for autocatalytic  chemical  reactions  \cite{schlogl,delatorre},
and a  phenomenological  Euclidean  field theory of high energy hadronic
collision processes (Reggeon field theory)  \cite{delatorre,cardysugar}.
Further   investigation   revealed  that  the  DP   universality   class
accommodates  more  general  models,  with more than a single  component
\cite{zgb,grinstein,fogedby}  as well as with  multiple,  in some  cases
infinitely              many              absorbing               states
\cite{pcp,jensendick,albano,munoz,janssen}.    Actually,    even    some
non-equilibrium  models  without  absorbing  states at all were found to
share some of the DP exponents \cite{kertesz,alon,jrgm,mario}.

It has been found, however, that not all phase transitions  involving an
absorbing state fall into the DP universality class.  Examples of models
exhibiting non-DP critical  behavior range from  probabilistic  cellular
automata   \cite{dertwer}  and  two-temperature   kinetic  Ising  models
\cite{menyhard},   to   interacting    monomer-dimer    \cite{imd}   and
monomer-monomer models \cite{browne},  branching and annihilating random
walks with even number of offspring \cite{takayasu,jensen,zhong,tauber},
and a class of  parity-conserving  contact processes that is the subject
of this paper  \cite{cpm}.  The common  feature  of all these  models is
that the number of  interacting  particles,  whether  they are  actually
particles  or are domain  walls, is  locally  conserved  modulo 2, i.e.,
their local dynamical rules conserve  parity.  It thus appeared at first
that local  conservation  laws were  affecting the critical  behavior of
non-equilibrium  systems, as one may have guessed from his  knowledge of
equilibrium critical phenomena.  The new universality class that emerged
became  known  as  the  parity  conserving  (PC)   universality   class.
Controversy  arose  when some  parity-conserving  models  were  shown to
belong to the DP  universality  class.  In  \cite{park},  it was noticed
that when one adds a parity-conserving external field to the dynamics of
a certain  monomer-dimer model with two equivalent  absorbing states for
which the number of domain walls is conserved modulo 2, the universality
class of the model crosses from the PC class to the DP class.  This kind
of   behavior   was  then   subsequently   observed   in  other   models
\cite{haye,hwang},  and  nowadays  it is  believed  that  it is not  the
symmetry of parity conservation that determines the critical behavior in
these  systems,  but  that  it  is  the  presence  of  two   equivalent,
$Z_{2}$-symmetric  absorbing  states  that  matters.  This  has led some
people  to refer to the new  universality  class as the  directed  Ising
universality class, in allusion to the fact that the Ising model has two
$Z_{2}$-symmetric  equivalent ground states.  Whether the PC class is as
robust as the DP class remains  largely an open  question,  and there is
lot of room left to research; for a review, see \cite{marro,advphys}.

In this paper we provide finite-size scaling estimates for the dynamical
critical exponent  $z=\nu_{\|}/\nu_{\perp}$ of the PC universality class
of critical  behavior  through exact numerical  diagonalizations  of the
master operator of a certain  parity-conserving  contact  process in one
dimension  \cite{cpm}.  We show that upon the  introduction of diffusion
in the model its critical behavior crosses over to another  universality
class that seem to be characterized  by DP exponents,  although our data
are not very good in this case.  We also present a brief  discussion  of
the  many-sector  decomposability  of the  state  space  of a  class  of
parity-conserving  contact  processes,  and how  diffusion  breaks  this
structure,  making the models with diffusion  strong  candidates to show
new types of dynamical critical behavior.

\section{Parity-conserving contact processes with diffusion}

Contact  processes may be viewed as models for the spread of an epidemic
among individuals  living in a  $d$-dimensional  lattice.  We consider a
parity-conserving  generalization  of the basic  contact  process  first
introduced in \cite{cpm}.  In this class of processes,  called  CP($m$),
an array of $m$ adjacent healthy individuals  $\emptyset$  surrounded by
$k$ infected  individuals $X$ becomes infected at rate $k\lambda$, while
an array of $m$ adjacent  infected  individuals  becomes healthy at unit
rate.  Pictorially,  in one  dimension we have for CP(1) the  elementary
processes $X \emptyset \emptyset \stackrel{\lambda}{\to} X X \emptyset$,
$\emptyset  \emptyset  X  \stackrel{\lambda}{\to}  \emptyset  X  X$,  $X
\emptyset X  \stackrel{2\lambda}{\to}  X X X$, and $X  \stackrel{1}{\to}
\emptyset$,  corresponding to the usual basic contact process, while for
CP(2) we have the elementary  processes $X \emptyset \emptyset \emptyset
\stackrel{\lambda}{\to} X X X \emptyset$, $\emptyset \emptyset \emptyset
X  \stackrel{\lambda}{\to}  \emptyset X X X$, $X  \emptyset  \emptyset X
\stackrel{2\lambda}{\to} X X X X$, and $X X \stackrel{1}{\to}  \emptyset
\emptyset$,  and  analogously  for $m>2$.  We clearly  see that  CP($m$)
processes conserve the number of particles modulo $m$.

For finite systems with an absorbing  state, the steady state  coincides
with it.  For the CP(1)  process, it is simply  given by the  completely
empty lattice ${\bf  0}=(0,0,\ldots,0)$.  CP(2) has, besides  ${\bf 0}$,
the           two           other            absorbing            states
$(\emptyset,X,\emptyset,X,\ldots,\emptyset,X)$                       and
$(X,\emptyset,X,\emptyset,\ldots,X,\emptyset)$.  Under periodic boundary
conditions  these  two  configurations  are  the  same;  let us  call it
$\tilde{\bf  0}$.  Whether  $\tilde{\bf  0}$  belongs  to the  space  of
allowed configurations depends on the parity of the lattice size $L$ and
on the parity  sector $N \bmod 2$, where $N$ is the number of  particles
of the  initial  configuration.  If $L$ is odd, then the only  absorbing
state for CP(2) is ${\bf 0}$, since it is impossible to fill the lattice
with a repetition of  $\emptyset  X$'s only.  For a lattice of even size
$L$, the state  $\tilde{\bf 0}$ has $L/2$ particles, and this number has
to be compatible with the parity of $N \bmod 2$:  if $L=4l$, with $l \in
{\Bbb N}$, and $N \bmod 2=0$, then $\tilde{\bf 0}$ is an absorbing state
of the  process;  if  otherwise  $L=4l+2$,  then  $\tilde{\bf  0}$ is an
absorbing  state only if $N \bmod 2=1$.  In this work, for  definiteness
we carried out our  calculations  on lattices of odd size $L$ and on the
$N \bmod 2 = 0$ parity  sector,  such that the only steady  state of the
process is ${\bf 0}$.  Our  methods  do of course  equally  apply to the
other sectors of the dynamics as well.

The existence of more than one absorbing state for CP(2) already signals
a general  feature of CP($m$)  processes,  that the number of  absorbing
states of these  models  grows  with $m$.  Actually,  for $m \geq 3$ the
number of absorbing states grows exponentially with the system size $L$,
a property called many-sector  decomposability first verified in a class
of   adsorption-desorption   processes   of  $m$-mers   on  the  lattice
\cite{jammed,barma}.  The number  $I_{m}(L)$ of absorbing,  fully jammed
states  in  these  systems  grows   asymptotically   as  $I_{m}(L)  \sim
2\phi^{L}$,  where  $\phi$  is the  largest  real  root of  $\phi^{m}  =
2\phi^{m-1}-1$; for $m=3$, $\phi =\frac{1}{2}(1+\sqrt{5}) \simeq 1.618$,
the golden  mean, while for $m=4$,  $\phi  \simeq  1.839$.  Notice  that
$\phi < 2$ for  all $m <  \infty$.  This  many-sector  structure  of the
phase space can,  however, be broken by adding  diffusion  $X  \emptyset
\rightleftharpoons \emptyset X$ at a finite rate $\mu$ to the processes,
since it allows  the  otherwise  jammed  states  to  evolve,  eventually
leading to a state with an array of $m$  adjacent  healthy  or  infected
sites that can then react  according  to the  CP($m$)  rules.  Diffusion
thus reduces the number of absorbing  states of CP($m$)  from  $I_{m}(L)
\sim  2\phi^{L}$  to  $I_{m}(L)=m$,   corresponding  to  the  number  of
$m$-parity  equivalent  absorbing  states.  While it is well  known that
diffusion,  as long as the  diffusion  constant  remains  finite,  is an
irrelevant   perturbation   for  the   basic   contact   process   CP(1)
\cite{irrelevant},  for  systems  with  more  than one  absorbing  state
diffusion  may  become a  highly  relevant  perturbation,  changing  the
critical  behavior of the  process,  as  recently  verified  in the pair
contact process with diffusion \cite{carlon,diff}.

\section{Finite-size scaling}

As is well  known  \cite{felder,siggia,adhr},  we may write  the  master
equation  for   reaction-diffusion   processes   on  the  lattice  as  a
Schr\"{o}dinger-like equation in Euclidean time,
\begin{equation}
\frac{d}{dt}|P(t)\rangle = -H|P(t)\rangle,
\end{equation}
with  $|P(t)\rangle$ the generating vector of the probabilities  $P({\bf
n},t)=\langle{\bf    n}|P(t)\rangle$   of   observing   the   particular
configuration       ${\bf        n}=(n_{1},n_{2},\ldots,n_{L})       \in
\{\emptyset,X\}^{L}$  at  instant  $t$, and with  $H$ the  infinitesimal
generator     of    the    Markov     semigroup.    The    lowest    gap
$E_{L}^{(1)}-E_{L}^{(0)}=E_{L}^{(1)}$ in the spectrum of $H$ may be used
to perform a finite-size scaling analysis in the same way as one does in
equilibrium  problems   \cite{kinzel,fss}.  Around  the  critical  point
$\lambda \gtrsim  \lambda^{*}$, the correlation  lengths of the infinite
system behave like
\begin{equation}
\label{CORR}
\xi_{\|} \propto \xi_{\perp}^{z} \propto (\lambda-\lambda^{*})^{-\nu_{\|}}
                                 \propto (\lambda-\lambda^{*})^{-\nu_{\perp}z},
\end{equation}
where  $\xi_{\|}$  and   $\xi_{\perp}$   are  the  correlation   lengths
respectively   in  the  time  and  space   directions,   $\nu_{\|}$  and
$\nu_{\perp}$   are   the   corresponding    critical   exponents,   and
$z=\nu_{\|}/\nu_{\perp}$ is the dynamical critical exponent.  For finite
systems of size $L$, we expect that
\begin{equation}
\label{CSI}
\xi_{\|,L}^{-1} = 
L^{-z_{L}}\Phi\left(|\lambda-\lambda^{*}_{L}|L^{1/\nu_{\perp,L}}\right),
\end{equation}
where  $z_{L}$ and  $\nu_{\perp,L}$  are the finite  versions of $z$ and
$\nu_{\perp}$,  and $\Phi(u)$ is a scaling  function with $\Phi(u \gg 1)
\sim u^{\nu_{\|}}$.  On general grounds one expects $\lim_{L \to \infty}
p^{*}_{L},z_{L},\nu_{\perp,L}=p^{*},      z,      \nu_{\perp}$.     From
Eqs.~(\ref{CORR}) and (\ref{CSI}) we obtain
\begin{equation}
\label{PCTHETA}
      \frac{\ln \left[ \xi_{\|,L}(\lambda^{*}_{L})/\xi_{\|,L'}(\lambda^{*}_{L})
               \right]}{\ln (L/L')} =
      \frac{\ln\left[ \xi_{\|,L''}(\lambda^{*}_{L})/\xi_{\|,L}(\lambda^{*}_{L})
                \right]}{\ln(L''/L)}=
      z_{L},
\end{equation}
which through the comparison of three different system sizes  $L'<L<L''$
furnishes  simultaneously  $\lambda^{*}_{L}$  and  $z_{L}$.  Of  course,
$\xi_{\|,L}$   and  the  gap   $E_{L}^{(1)}$   of  $H$  are  related  by
$\xi_{\|,L}^{-1}= {\rm Re}\{ E_{L}^{(1)}\}$.

\section{Dynamical critical exponents}

We calculated  the gaps of $H$ through the power method, which  requires
only   matrix-by-vector   multiplications   that  can  be  carried   out
efficiently, and does not require a  diagonalization  in the usual, `QR'
sense, a step that may lessen the  quality of the data.  The  particular
implementation  of the power method we use takes full  advantage  of the
presence of absorbing  states in the process,  and is also  suitable for
the  investigation of time dependent  properties of finite-state  Markov
chains \cite{power}.

\begin{figure}[t]
\centerline{\epsfig{figure=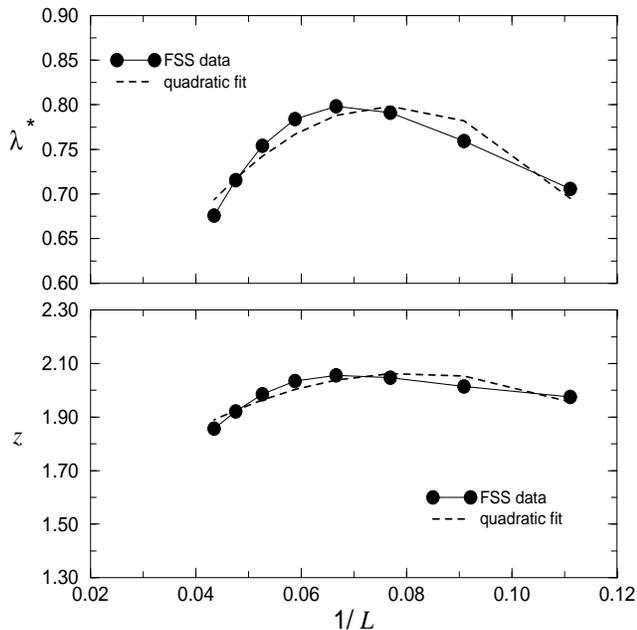,height=86mm,width=86mm}}
\caption{Finite-size  data and least-squares fits for the critical point
$\lambda^{*}$ and the dynamical critical exponent $z$ of the CP(2) model
with diffusion.}
\label{FIG}
\end{figure}

Our  results  for  $\lambda^{*}$  and $z$ for  the  CP(2)  both  without
diffusion and with symmetric  diffusion $X \emptyset  \rightleftharpoons
\emptyset X$ at rate  $\mu=0.05$  are  summarized in Table  \ref{TABLE}.
The choice  $\mu=0.05$ is arbitrary,  except that it  represents  only a
small  perturbation  to the main  coupling  $\lambda$  near the critical
points,  but not so small as to  render  too  much  unbalanced,  `stiff'
matrices  that may  become  difficult  to  diagonalize.  We were able to
diagonalize  systems with up to $L=25$ sites in a  reasonable  amount of
time   and   memory   space,   comparable   to   recent   density-matrix
renormalization  group  studies of similar  systems  \cite{carlon}.  The
extrapolated  values  for the  $\mu=0$  case  were  obtained  through  a
Bulirsch-Stoer  extrapolation  \cite{bst}, while those for the $\mu \neq
0$  case  were  obtained  through  a  least-squares  fit  to  the  curve
$y_{L}=a_{0}+    a_{1}L^{-1}+a_{2}L^{-2}$,    and   can   be   seen   in
Fig.~\ref{FIG}.  The  uncertainties  associated  with  our  extrapolated
numbers  are  mainly  due to  finite-size  effects  and  corrections  to
scaling, as well as to the extrapolation procedures itself.  We obtained
in the  diffusionless  case  $\lambda^{*}=0.88  \pm 0.01$  ($\omega_{\rm
BST}=0.924$) and $z=1.75 \pm 0.01$  ($\omega_{\rm  BST}=1.802$), and for
the diffusive case  $\lambda^{*}=0.25 \pm 0.05$ and $z=1.3 \pm 0.1$.  (A
linear fit $y_{L}=a_{0}+a_{1}L^{-1}$ to the last four points of the data
with diffusion furnishes  $\lambda^{*}=0.38 \pm 0.04$ with a correlation
coefficient  $\gamma=0.989$,  and $z=1.36  \pm 0.06$ with a  correlation
coefficient  $\gamma=0.989$.)  We thus see that the pure  CP(2)  without
diffusion has a critical  behavior  governed by the  dynamical  critical
exponent of the parity-conserving universality class, for which the best
known  value  to  date  is  given  by  $z_{\rm   PC}=1.750  \pm  0.005)$
\cite{jensen}.  The value of the  critical  point  agrees  well with the
value  $\lambda^{*}=0.8935  \pm 0.0004$  found in  \cite{cpm}.  Upon the
introduction of diffusion, however, the number of steady states of CP(2)
reduces from two to one, and according to our data its critical behavior
crosses over to that of another  universality  class.  In a first moment
the critical  exponent seem to be converging  to $z \simeq 2$.  Data for
larger  lattice  sizes,  however,  point  toward a lower  value  of $z$,
definitely different from $z=2$, at least as far as our finite-size data
go.  We  also  observed  a  non-monotonic  behavior  of the  data,  that
unfortunately  seems to be more common in this type of calculation  than
it would be desirable.  This non-monotonic  behavior is probably related
with  the  existence  of  a  whole   critical  (in  the  present   case,
diffusion-like)  phase  to the  right  of the  critical  point,  and has
already been observed in other studies of systems with extended critical
phases  \cite{jrgm,carlon}.  Our  extrapolation  gives  $z=1.3 \pm 0.1$,
closer to the exponent of the DP universality  class, namely $z_{\rm DP}
\simeq 1.5807$  \cite{power,precise},  than to other known values.  A DP
critical behavior is what one could have expected on the basis of the DP
conjecture,  since the process  with  diffusion  has a single  absorbing
state.  Intuition,  however,  has proved not to be reliable  in guessing
the critical behavior of nonequilibrium  systems with local conservation
laws or many absorbing  states, as one infers from the recent history in
the field, and the presence of an  additional  symmetry  might well have
driven the critical exponents of CP(2) with diffusion to those of the PC
universality  class.  Diffusion  is then  seen to be a  highly  relevant
perturbation for systems with more than one absorbing state.

\begin{table}[t]
\centering
\caption{Finite-size data and extrapolated values for the critical point
and the  dynamical  critical  exponent  $z$ of CP(2) with  diffusion  $X
\emptyset  \rightleftharpoons  \emptyset  X$ at rate $\mu$.  The numbers
between parentheses  represent the estimated errors in the last digit of
the data.}
\label{TABLE}
\begin{tabular}{ccccc}
System sizes & \multicolumn{2}{c}{$\mu=0$} & \multicolumn{2}{c}{$\mu=0.05$} \\
$L',L,L''$   & $\lambda^{*}_{L}$ & $z_{L}$ & $\lambda^{*}_{L}$ & $z_{L}$    \\
\hline
7,9,11         & 0.584\,505
               & 1.916\,330
               & 0.705\,564
               & 1.975\,081    \\
9,11,13        & 0.607\,946
               & 1.898\,353
               & 0.759\,294
               & 2.013\,974    \\
11,13,15       & 0.625\,558
               & 1.884\,819
               & 0.791\,092
               & 2.046\,900    \\
13,15,17       & 0.641\,431
               & 1.872\,620
               & 0.798\,318
               & 2.056\,315    \\
15,17,19       & 0.656\,023
               & 1.861\,476
               & 0.783\,969
               & 2.034\,662    \\
17,19,21       & 0.669\,350
               & 1.851\,398
               & 0.753\,899
               & 1.985\,477    \\
19,21,23       & 0.681\,446
               & 1.842\,353
               & 0.715\,444
               & 1.921\,154    \\
21,23,25       & 0.692\,390
               & 1.834\,263
               & 0.675\,905
               & 1.856\,800    \\
\hline
Extrapolated   & 0.88(1)
               & 1.75(1)
               & 0.25(5)
               & 1.3(1)        \\
\end{tabular}
\end{table}

\section{Summary and conclusions}

In summary, we have  conducted a  finite-size  scaling  analysis  of the
dynamical critical exponent of an even-parity-conserving contact process
through exact numerical diagonalizations of its time evolution operator.
We showed  that the  critical  behavior  of the  `pure'  model  with two
absorbing  configurations belongs to the parity-conserving  universality
class.  In the presence of symmetric  diffusion,  however, the number of
absorbing  configurations  in the model reduces from two to one, and its
critical  behavior  crosses over to that of another  universality  class
that seem to be that of the directed percolation process.  This behavior
is in  accordance  with what one  expects  on the basis of the  directed
percolation conjecture, at least for small diffusion rates.

It  would  be  interesting   to  perform   time-dependent   Monte  Carlo
simulations of the CP(2) model both with and without  diffusion in order
to  determine  its  critical  exponents  more  precisely,  as well as to
investigate the critical  behavior of other members of the CP($m$) class
of  processes.  In  particular,   it  would  be  very   interesting   to
investigate  the CP(3),  since it was argued on the basis of Monte Carlo
simulations  and Pad\'{e}  approximants  analyses that this process does
not  suffer  an  ordinary  second  order  phase  transition  \cite{cpm}.
Moreover,  the  CP(3)  process  has  the   many-sector   decomposability
property, and it would be interesting  to see how diffusion  changes the
scenario   for  the   phase   transitions,   if  any,  in  this   model.

It seems that the question as to the roles of symmetries and many-sector
decomposability  on the critical behavior of nonequilibrium  interacting
particle  systems is far from being  answered,  and more  numerical  and
analytical work has to be done before a consistent scenario emerges.

\section*{ACKNOWLEDGMENT}

The author would like to acknowledge  Professor Pablo A.  Ferrari of the
Instituto de  Matem\'{a}tica e  Estat\'{\i}stica  of the Universidade de
S\~{a}o  Paulo,  Brazil, for his kind  hospitality  while the author was
there.  The author also  acknowledges the Departamento de F\'{\i}sica of
the Universidade  Federal de S\~{a}o Carlos, Brazil, for the use of part
of its computational resources, and the Funda\c{c}\~{a}o de Amparo \`{a}
Pesquisa  do Estado de  S\~{a}o  Paulo  (FAPESP),  Brazil,  for  partial
financial support.


\begin{references}

\bibitem[*]{email}E-mail: jricardo.mendonca@edb.ericsson.se.

\bibitem{zpb42}H.K. Janssen, Z. Phys. B {\bf 42}, 151 (1981).

\bibitem{zpb47}P. Grassberger, Z. Phys. B {\bf 47}, 365 (1982).

\bibitem{kinzel}W. Kinzel and J.M. Yeomans, J. Phys. A {\bf 14},
L162 (1981); W. Kinzel, Z. Phys. B {\bf 58}, 229 (1985).

\bibitem{harris}T.E. Harris, Ann. Prob. {\bf 2}, 969 (1974).

\bibitem{schlogl}F. Schl\"{o}gl, Z. Phys. {\bf 253}, 147 (1972).

\bibitem{delatorre}P. Grassberger and A. de la Torre, Ann. Phys.
(N.Y.) {\bf 122}, 373 (1979).

\bibitem{cardysugar}J.L. Cardy and R.L. Sugar, J. Phys. A {\bf 13},
L423 (1980).

\bibitem{zgb}R.M. Ziff, E. Gulari, and Y. Barshad, Phys. Rev. Lett.
{\bf 56}, 2553 (1986).

\bibitem{grinstein}G. Grinstein, Z.-W. Lai, and D.A. Browne, Phys. Rev.
A {\bf 40}, 4820 (1989).

\bibitem{fogedby}I. Jensen, H.C. Fogedby, and R. Dickman, Phys. Rev.
A {\bf 41}, 3411 (1990).

\bibitem{pcp}I. Jensen, Phys. Rev. Lett. {\bf 70}, 1465 (1993).

\bibitem{jensendick}I. Jensen and R. Dickman, Phys. Rev. E {\bf 48}, 
1710 (1993).

\bibitem{albano}E.V. Albano, Physica A {\bf 214}, 426 (1995).

\bibitem{munoz}M.A. Mu\~{n}oz, G. Grinstein, R. Dickman, and R. Livi,
Phys. Rev. Lett. {\bf 76}, 451 (1996); M.A. Mu\~{n}oz, G. Grinstein,
and R. Dickman, J. Stat. Phys. {\bf 91}, 541 (1998).

\bibitem{janssen}H.K. Janssen, Phys. Rev. Lett. {\bf 78}, 2890 (1997).

\bibitem{kertesz}J. Kert\'{e}sz and D.E. Wolf, Phys. Rev. Lett. {\bf 62},
2571 (1989).

\bibitem{alon}U. Alon, M.R. Evans, H. Hinrichsen, and D. Mukamel,
Phys. Rev. Lett. {\bf 76}, 2746 (1996); Phys. Rev. E {\bf 57}, 4997 (1998).

\bibitem{jrgm}J.R.G. de Mendon\c{c}a, Phys. Rev. E {\bf 60}, 1329 (1999).

\bibitem{mario}T. Tom\'{e} and M.J. de Oliveira, Phys. Rev. Lett. {\bf 86},
5643 (2001).

\bibitem{dertwer}P. Grassberger, F. Krause, and T. von der Twer, J. Phys.
A {\bf 17}, L105 (1983); P. Grassberger, J. Phys. A {\bf 22}, L1103 (1989).

\bibitem{menyhard}N. Menyh\'{a}rd, J. Phys. A {\bf 27}, 6139 (1994);
N. Menyh\'{a}rd and G. \'{O}dor, J. Phys. A {\bf 28}, 4505 (1995).

\bibitem{imd}M.H. Kim and H. Park, Phys. Rev. Lett. {\bf 73}, 2579 (1994);
H. Park, M.H. Kim, and H. Park, Phys. Rev. E {\bf 52}, 5664 (1995).

\bibitem{browne}K.E. Bassler and D.A. Browne, Phys. Rev. Lett. {\bf 77},
4094 (1996); Phys. Rev. E {\bf 55}, 5225 (1997).

\bibitem{takayasu}H. Takayasu and A.Yu. Tretyakov, Phys. Rev. Lett.
{\bf 68}, 3060 (1992).

\bibitem{jensen}I. Jensen, Phys. Rev. E {\bf 50}, 3623 (1994).

\bibitem{zhong}D. Zhong and D. ben-Avraham, Phys. Lett. A {\bf 209}, 
333 (1995).

\bibitem{tauber}J. Cardy and U.C. T\"{a}uber, Phys. Rev. Lett. {\bf 77},
4780 (1996); J. Stat. Phys. {\bf 90}, 1 (1998).

\bibitem{cpm}N. Inui and A.Yu. Tretyakov, Phys. Rev. Lett. {\bf 80}, 
5148 (1998).

\bibitem{park}H. Park and H. Park, Physica A {\bf 221}, 97 (1995).

\bibitem{haye}H. Hinrichsen, Phys. Rev. E {\bf 55}, 219 (1997).

\bibitem{hwang}W. Hwang, S. Kwon, H. Park, and H. Park, Phys. Rev.
E {\bf 57}, 6438 (1998).

\bibitem{marro}J. Marro and R. Dickman, {\it Nonequilibrium Phase Transitions
in Lattice Models\/} (Cambridge University Press, Cambridge, 1999).

\bibitem{advphys}H. Hinrichsen, Adv. Phys. {\bf 49}, 815 (2000).

\bibitem{jammed}M. Barma, M.D. Grynberg, and R.B. Stinchcombe, Phys. Rev.
Lett. {\bf 70}, 1033 (1993); R.B. Stinchcombe, M.D. Grynberg, and M. Barma,
Phys. Rev. E {\bf 47}, 4018 (1993).

\bibitem{barma}M. Barma, in {\it Nonequilibrium Statistical Mechanics
in One Dimension\/}, edited by V. Privman (Cambridge University Press,
Cambridge, 1997), and references there in.

\bibitem{irrelevant}R. Dickman, Phys. Rev. B {\bf 40}, 7005 (1989);
I. Jensen and R. Dickman, J. Phys. A {\bf 26}, L151 (1993).

\bibitem{carlon}E. Carlon, M. Henkel, and U. Schollw\"{o}ck,
e-print cond-mat/9912347.

\bibitem{diff}G. \'{O}dor,
Phys. Rev. E {\bf 62}, R3027 (2000);
H. Hinrichsen, Phys. Rev. E {\bf 63}, 36102 (2001).

\bibitem{felder}B.U. Felderhof, Rep. Math. Phys. {\bf 1}, 215 (1971);
{\bf 2}, 151 (1971).

\bibitem{siggia}E.D. Siggia, Phys. Rev. B {\bf 16}, 2319 (1977).

\bibitem{adhr}F.C. Alcaraz, M. Droz, M. Henkel, and V. Rittenberg,
Ann. Phys. (N.Y.) {\bf 230}, 250 (1994).

\bibitem{fss}{\it Finite-size Scaling and Numerical Simulations of Statistical
Systems\/}, edited by V. Privman (World Scientific, Singapore, 1990).

\bibitem{power}J.R.G. de Mendon\c{c}a, J. Phys. A {\bf 32}, L467 (1999).

\bibitem{bst}R. Bulirsch and J. Stoer, Numer. Math. {\bf 6}, 413 (1964);
M. Henkel and G. Sch\"{u}tz, J. Phys. A {\bf 21}, 2617 (1988).

\bibitem{precise}I. Jensen, J. Phys. A {\bf 32}, 5233 (1999).

\end{references}
\end{document}